%% file: main.tex
\documentclass[11pt]{article}

\usepackage[margin=1in]{geometry}
\usepackage[T1]{fontenc}
\usepackage{lmodern}
\usepackage{microtype}
\usepackage{amsmath,amssymb}
\usepackage{graphicx}
\usepackage{booktabs}
\usepackage{multirow}
\usepackage{threeparttable}
\usepackage{caption}
\usepackage{float}
\usepackage{placeins}
\usepackage{xcolor}
\usepackage{authblk}
\usepackage[numbers,sort&compress]{natbib}
\usepackage[colorlinks=true,linkcolor=blue!60!black,citecolor=blue!60!black,urlcolor=blue!60!black]{hyperref}
\newcommand{\doi}[1]{\href{https://doi.org/#1}{doi:\nolinkurl{#1}}}

\graphicspath{{figures/}}
\newcommand{\OT}{Omic Translator}
\newcommand{\rhoG}{\rho_{\mathrm{gene}}}
\newcommand{\rhoC}{\rho_{\mathrm{cell}}}

\input{generated/numbers.tex}

\title{\textbf{An autoencoder outperforms two transformers on prediction of gene
expression from chromatin accessibility}}

\author[1]{Jesse G. Meyer}
\affil[1]{Department of Computational Biomedicine, Cedars-Sinai Medical Center, Los Angeles, CA, USA}
\date{Draft of \today}

\begin{document}
\maketitle

\begin{abstract}
\noindent
Models that predict RNA abundance from chromatin accessibility are rarely compared under
the same features, targets and metrics. We built one benchmark on 126{,}604 paired
single-nucleus profiles from 19 human studies with a shared vocabulary of 50{,}000 peaks
and 3{,}185 genes, ten $5{\times}2$ cross-validation folds and one scoring target, and
compared seven implemented pipelines and two reference predictors across 70 trained
runs; each pipeline keeps its own input normalisation and native training target. The scButterfly-B pipeline, built on dual variational autoencoders,
outperformed every transformer pipeline, including an autoregressive Perceiver model
(\OT{}) and a support-free reformulation of Cisformer, on per-gene and within-cell
Spearman correlation in all sixty paired fold comparisons, with \speedRatioPrimary{}
times shorter training. Transformer pipelines come within \gapWcPercMin--\gapWcPercMax{}
on within-cell profile shape but lag by \gapPgMin--\gapPgMax{} on per-gene correlation.
On two withheld cell types per-gene correlation is low; controls show the podocyte values
are typical of within-type correlation for seen cell types of the same size, whereas
Paneth values match a cell-agnostic shuffle for every \OT{} arm. Within the transformer
family, prespecified ablations show the Perceiver encoder was worth
$\deltaPerceiverCompressionPg$ in within-donor per-gene correlation over full
self-attention, a chromosome-aware genomic attention prior reduced it by
\adeltaGenomicSpatialPriorPg{} (exact donor-level sign-flip test, Holm-adjusted
$p = \pholmGenomicSpatialPriorPg$), and decoder output order had no detectable effect.
A training-mean profile reaches \wcMean{} within-cell correlation with no input and a
constant at the true expressed-gene positions reaches \pgOracle{} per-gene correlation,
so oracle-support protocols are reported only as diagnostics. Code, results and an
audited release are provided.
\end{abstract}

\section{Introduction}

Paired single-cell measurements of chromatin accessibility and RNA abundance offer a
direct supervised signal for learning how regulatory state maps to transcriptional
output. A model that learned this mapping well could impute an unmeasured modality,
denoise sparse measurements, and, if its internal structure were interpretable, point at
regulatory elements that matter for particular genes. A number of architectures have been
proposed for the translation task. BABEL couples modality-specific autoencoders through a
shared latent space~\citep{wu2021babel}; scButterfly aligns two variational autoencoders
with discriminators and supports translation in both directions~\citep{cao2024scbutterfly};
Polarbear trains semi-supervised autoencoders on partially paired
data~\citep{zhang2022polarbear}; CMOT frames translation as optimal
transport~\citep{alatkar2023cmot}; MultiVI and GLUE learn joint embeddings that can be
decoded into either modality~\citep{ashuach2023multivi,cao2022glue}. More recently,
attention-based models such as Cisformer~\citep{ji2025cisformer} and foundation models
pre-trained on large ATAC compendia~\citep{wu2025epifoundation,liu2025scarf} have been
applied to the same problem.

Two things make it difficult to learn from these papers collectively. The first is that
they rarely share a feature space, a split, or a metric, so a reported improvement can
come from any of those choices rather than from the architecture. The second is that
several of them supply information at inference time that a deployed imputation model
would not have. Cisformer, for example, is evaluated by asking the decoder for abundance
values only at genes already known from the target RNA profile to be expressed. That is a
legitimate task, but it is a different and much easier task than generating an RNA
profile from chromatin alone: we show below that a constant written at the true
expressed-gene positions scores \pgOracle{} on per-gene correlation without any model.
Recent benchmarking of single-cell deep learning in other settings has found that simple
baselines are often competitive with or better than large
models~\citep{ahlmanneltze2025,kedzierska2025zeroshot}, which makes it important to
include deliberately trivial controls rather than only published competitors.

This paper has two aims. The first is to test, under prespecified contrasts with
seeds, batch size, peak cap and hardware held fixed, which parts of a transformer
translator actually contribute to accuracy. The architecture we test, \OT{}, was
designed around three ideas that seemed sensible a priori: that compressing thousands of
sparse peak events into a small set of latents is better than attending over all of
them~\citep{jaegle2021perceiver}; that attention from a latent to a peak should be biased
toward peaks on the same chromosome and penalised with genomic distance, in the spirit of
linear attention biases~\citep{press2022alibi}; and that an autoregressive decoder should
emit genes in a fixed order so that earlier emissions can condition later ones. Only the
first of these survived contact with the data.

The second aim is to place that model family against published pipelines in a way that
is honest about what the comparison can and cannot support. We compare the implemented
\OT{}, scButterfly-B and support-free Cisformer pipelines, each run with its own input
normalisation and native training target, against one common scoring target on
126{,}604 paired nuclei from 19 human studies, with ten $5{\times}2$ cross-validation
folds~\citep{dietterich1998}, every metric recomputed from saved cell-level
predictions, inference clustered on the donor, and two reference predictors that bracket
the metric scale. The result is a clear ranking of pipelines that we decline to
generalise into a statement about architecture classes or about models trained on
equivalent targets, for reasons we spell out.

\section{Results}

\FloatBarrier
\subsection{A benchmark in which only the model changes}

\begin{figure}[!htb]
  \centering
  \includegraphics[width=0.95\textwidth]{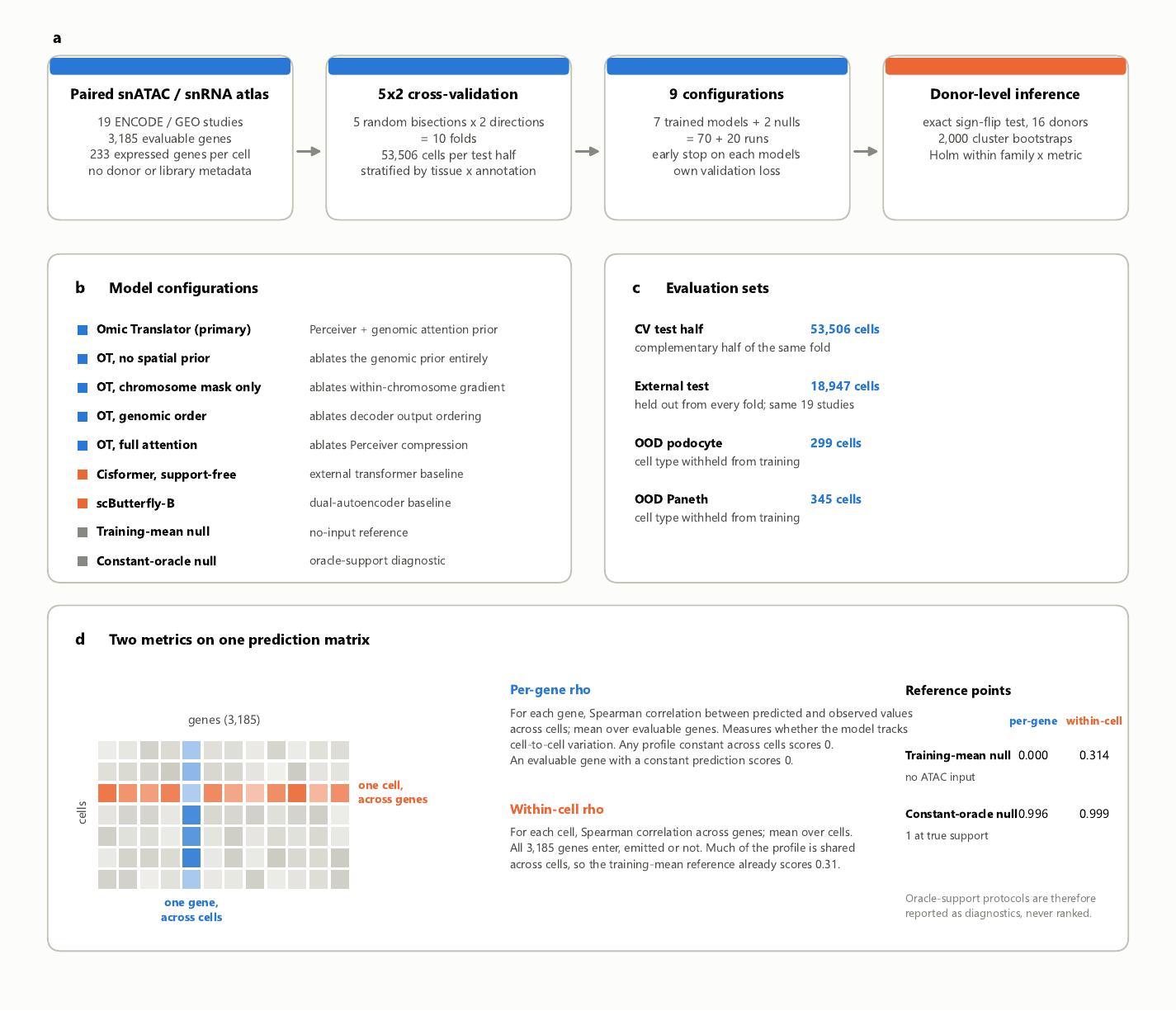}
  \caption{\textbf{Experimental design and metrics.} (a) Paired snATAC/snRNA profiles
  from 19 studies are partitioned by $5{\times}2$ cross-validation into ten folds; every
  model early-stops on its own validation loss. (b) The seven trained configurations and
  two reference predictors. (c) The four evaluation sets. (d) The two metrics computed on
  one prediction matrix: per-gene correlation runs down a column (one gene, across cells)
  and within-cell correlation runs along a row (one cell, across all 3{,}185 genes,
  emitted or not). The two reference predictors anchor each metric.}
  \label{fig:design}
\end{figure}

The experimental design is summarised in Figure~\ref{fig:design}. The data are 126{,}604
quality-controlled nuclei with paired snATAC-seq and snRNA-seq from the multi-tissue
atlas distributed with EpiFoundation~\citep{wu2025epifoundation}, which draws on 19
ENCODE~\citep{encode2020} and GEO studies, one study per tissue (Methods;
Supplementary Table~\ref{tab:studies}). All pipelines receive the same 50{,}000-peak
accessibility vocabulary and are scored against the same 3{,}185-gene target matrix with
one metric implementation. The feature vocabularies were selected once on the whole atlas
before any split was made, using the RNA values of every cell including those later held
out, so feature selection is transductive; each pipeline then applies its own input
normalisation and trains on its own native target (Methods). What is compared is the
implemented pipelines as run, scored on a common target. Two cell types, podocytes ($n = \nPodocyte$,
kidney cortex) and Paneth cells ($n = \nPaneth$, ileum), were removed from an initial
stratified partition and serve as withheld sets. The \nCellsExternal{} cells of that
partition's test fraction form a fixed same-atlas holdout that no run trains or validates
on. The remaining 107{,}013 cells form the cross-validation pool.

We use $5{\times}2$ cross-validation: five independent random bisections of the pool,
stratified by tissue and cell-type annotation, each used in both directions, giving ten
folds in which every cell in the pool is a test cell exactly five times. Within each
training half, 10\% is carved out as an inner validation set that is the only signal for
early stopping. The test half of each fold contains 53{,}506 or 53{,}507 cells and is
scored in full; we do not subsample cells for any metric. The ten folds are not ten
independent observations, and we never treat them as such; inference is described below.

The atlas object carries no donor, sample or library identifiers, and the per-study
accession is one-to-one with tissue, so group-protected splitting was not possible
without holding out whole tissues. We therefore split at the cell level and treat the
cross-validation estimates as in-distribution generalisation. Donor identity is, however,
recoverable from the upstream ENCODE and GEO records: the 11 ENCODE accessions come from
8 donors, one of whom contributes colon, adrenal gland and fallopian tube and another
muscle and bile duct, while each GEO study is a single sample (Supplementary
Table~\ref{tab:studies}). The independent sampling unit for inference is therefore the
donor, with 16 units rather than 19 studies, and the statistical analysis below is
clustered accordingly.

Seven trained pipelines enter the matrix: five variants of \OT{} that implement the
prespecified ablations, scButterfly-B, and Cisformer in a support-free mode. Two
reference predictors accompany them. The matrix is $7 \times 5 \times 2 = 70$ trained
runs plus 20 reference evaluations; every run completed, and every reported number is
recomputed from the saved prediction store.

\FloatBarrier
\subsection{Two metrics and two reference predictors}

We report two Spearman correlations (Figure~\ref{fig:design}d). The \emph{per-gene}
correlation $\rhoG$ asks, for each gene, whether the model orders cells correctly by that
gene's abundance, and then averages over evaluable genes, that is, genes whose target
varies across the scored cells (\evaluableCV{} of 3{,}185 on a cross-validation half).
A gene whose target varies but whose prediction is constant across cells receives
$\rhoG = 0$ rather than being dropped, so a sparse decoder cannot raise its average by
emitting only easy genes. \emph{Coverage} is the fraction of evaluable genes for which a
model's prediction varies across cells; it is a property of the gene axis, not a count
of genes emitted per cell. The \emph{within-cell} correlation $\rhoC$ asks, for each
cell, whether the model orders all 3{,}185 genes correctly within that cell, including
the genes it predicts as zero, and averages over cells.

The two metrics have different reference points, and we use two deliberately simple
predictors to mark them rather than treating either as a mathematical bound. A profile
equal to the training-mean expression, tiled across every test cell without looking at
chromatin, has $\rhoG = 0$ because every column is constant, but it reaches
$\rhoC = \wcMean$ because much of the within-cell rank structure is shared across cells.
A within-cell value near \wcMean{} is therefore consistent with no cell-specific
information, although matching that one scalar does not by itself show that a model
predicts the average cell. At the other end, a constant 1 written at exactly the
positions where the target is non-zero scores $\rhoG = \pgOracle$ and
$\rhoC = \wcOracle$. Any protocol that conditions on target support, including the
published Cisformer evaluation, must be read against that diagnostic, which is why we
evaluate Cisformer support-free and report oracle-support results only as diagnostics.

\FloatBarrier
\subsection{Genomic inductive biases reduce accuracy; the Perceiver encoder is what helps}

\begin{figure}[!htb]
  \centering
  \includegraphics[width=0.92\textwidth]{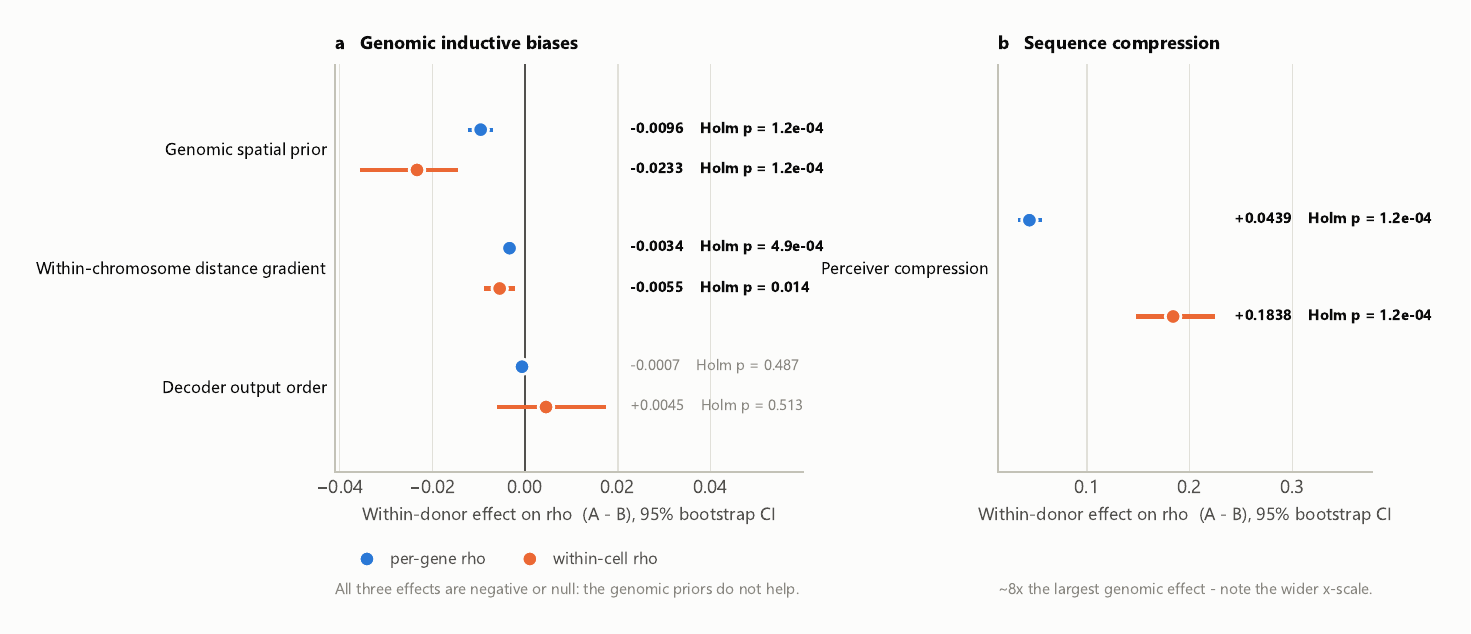}
  \caption{\textbf{Prespecified ablations.} Within-donor effect of each architectural
  contrast on per-gene (blue) and within-cell (orange) Spearman correlation, as the
  difference A $-$ B with 95\% donor-bootstrap intervals. (a) The three genomic inductive biases;
  (b) the Perceiver encoder against full self-attention, on a wider axis. Labels show the
  stored Holm-adjusted $p$-values; bold marks $p_{\mathrm{Holm}} < 0.05$.}
  \label{fig:ablations}
\end{figure}

The four prespecified contrasts are shown in Figure~\ref{fig:ablations} and
Table~\ref{tab:ablations}. Each compares two \OT{} variants that differ in one declared
setting, with model seed, batch size, peak cap, learning-rate schedule and GPU model
identical. Three of the contrasts change nothing but a mask or an ordering. The fourth,
Perceiver versus full attention, replaces the two cross-attention and two latent
self-attention blocks with two full self-attention blocks over the 8{,}192 tokens, gives
the decoder an 8{,}192-slot memory instead of 256 latents, and changes the parameter
count from 19.78 to 18.13 million; it is a comparison of those two implemented encoders,
not an isolation of compression alone.

\begin{table}[!htb]
  \centering
  \small
  \begin{threeparttable}
  \caption{\textbf{Prespecified architectural ablations on the cross-validation test
  half.} $\Delta$ is the mean over donors of the within-donor paired difference
  $\mathrm{A} - \mathrm{B}$ (each donor's difference being the mean over repetitions and
  directions of the metric computed on that donor's cells); it is smaller in magnitude
  than the pooled difference because between-study rank structure is excluded. $p$ is
  two-sided from \permDescAblations; Holm correction is
  within the four-contrast family and within metric. The 95\% interval is a percentile
  bootstrap over the same units (\nBootAblations{} draws). ``Reps'' counts repetitions
  whose sign agrees with $\Delta$.}
  \label{tab:ablations}
  \footnotesize
  \begin{tabular}{llrrrcc}
    \toprule
    Contrast (A vs.\ B) & Metric & $\Delta$ & $p$ & $p_{\mathrm{Holm}}$ & 95\% CI & Reps \\
    \midrule
    \input{generated/table_ablations.tex} \\
    \bottomrule
  \end{tabular}
  \begin{tablenotes}\footnotesize
    \item For Monte Carlo procedures a $p$-value of $1/(n+1)$ means that no draw matched or
    exceeded the observed statistic; it is an estimate, not an upper bound on the tail
    probability (Methods). Exact enumeration has no such floor.
  \end{tablenotes}
  \end{threeparttable}
\end{table}

The chromosome-aware genomic prior, which blocks each latent from attending to peaks on
other chromosomes and penalises attention within a chromosome linearly with distance in
megabases, hurt on both metrics. Removing it entirely improved $\rhoG$ by
\adeltaGenomicSpatialPriorPg{} and $\rhoC$ by \adeltaGenomicSpatialPriorWc{} (the
table reports primary minus none, so its entries are negative), with the same sign in
\repsGenomicSpatialPriorPg{} repetitions and intervals that exclude zero. Splitting the
prior into its two components shows that the within-chromosome distance gradient
accounts for about a third of the per-gene cost on its own
($\Delta = \deltaWithinChromosomeDistanceGradientPg$, $p_{\mathrm{Holm}} =
\pholmWithinChromosomeDistanceGradientPg$) and also lowers within-cell correlation
($\Delta = \deltaWithinChromosomeDistanceGradientWc$, $p_{\mathrm{Holm}} =
\pholmWithinChromosomeDistanceGradientWc$, interval
$[\ciLoWithinChromosomeDistanceGradientWc, \ciHiWithinChromosomeDistanceGradientWc]$).
In the pooled rank-swap sensitivity analysis the within-cell effect of the gradient is
not distinguishable from zero ($p = 0.35$); this is the only ablation contrast on which
the primary and the sensitivity analysis differ, and we report both.

The decoder's output order, which we had expected to matter because genomic order
matches the encoder's inductive bias, showed no detectable effect under this protocol.
Emitting genes in chromosome-then-TSS order or in one fixed random permutation gives
differences of $\deltaOutputOrderPg$ ($\rhoG$) and $\deltaOutputOrderWc$ ($\rhoC$) with
intervals of $[\ciLoOutputOrderPg, \ciHiOutputOrderPg]$ and
$[\ciLoOutputOrderWc, \ciHiOutputOrderWc]$; the sign varies across repetitions
(\repsOutputOrderPg{} and \repsOutputOrderWc{} agree with the mean). That is evidence
against effects outside those intervals, not evidence of exact equivalence.

Against this, the Perceiver encoder was worth $\deltaPerceiverCompressionPg$ on $\rhoG$
and $\deltaPerceiverCompressionWc$ on $\rhoC$ over the full-attention encoder,
\ratioPerceiverPg{} times the largest genomic-prior effect on per-gene correlation and
\ratioPerceiverWc{} times on within-cell correlation. The full-attention arm is also unstable:
its fold-to-fold standard deviation on $\rhoG$ is \pgFullSD, \fullSDratio{} times that of
any other arm, and its within-cell value of \wcFull{} sits close to the training-mean
reference. The two arms compared both have the spatial prior switched off, because the
full-attention encoder has no latents to anchor one to.

\FloatBarrier
\subsection{The autoencoder-based pipeline outperforms every transformer pipeline}

Table~\ref{tab:benchmark} gives the full ranking of the implemented pipelines on the cross-validation test half, scored on the common target.
scButterfly-B leads on both metrics, reaching
$\rhoG = \pgScb$ and $\rhoC = \wcScb$, with a mean end-to-end training time of
\trainHScb~h per fold on the same GPU against \trainHPrim~h for the primary \OT{} and
\trainHCis~h for Cisformer (Methods define what each timer covers). The best transformer
is the \OT{} variant with no spatial prior ($\rhoG = \pgNoPrior$, $\rhoC = \wcNoPrior$).
The primary \OT{} configuration, the one we would have reported had we not run the
ablations, sits at $\rhoG = \pgPrim$.

\begin{table}[!htb]
  \centering
  \small
  \begin{threeparttable}
  \caption{\textbf{Benchmark on the cross-validation test half.} Mean $\pm$ SD over ten
  folds of \nCellsTestHalf{} cells each. Coverage is the fraction of evaluable genes with a
  prediction that varies across cells; ``emitted'' is the mean number of non-zero
  predicted genes per cell against a target mean of \targetNonzero. Train time is the mean
  wall-clock per fold on one NVIDIA L40S, as timed by each runner. AE, autoencoder; TF,
  transformer.}
  \label{tab:benchmark}
  \footnotesize
  \begin{tabular}{llcccrr}
    \toprule
    Model & Type & $\rhoG$ & $\rhoC$ & Coverage & Emitted & Train (h) \\
    \midrule
    \input{generated/table_benchmark.tex} \\
    \bottomrule
  \end{tabular}
  \end{threeparttable}
\end{table}

\begin{figure}[!htb]
  \centering
  \includegraphics[width=0.9\textwidth]{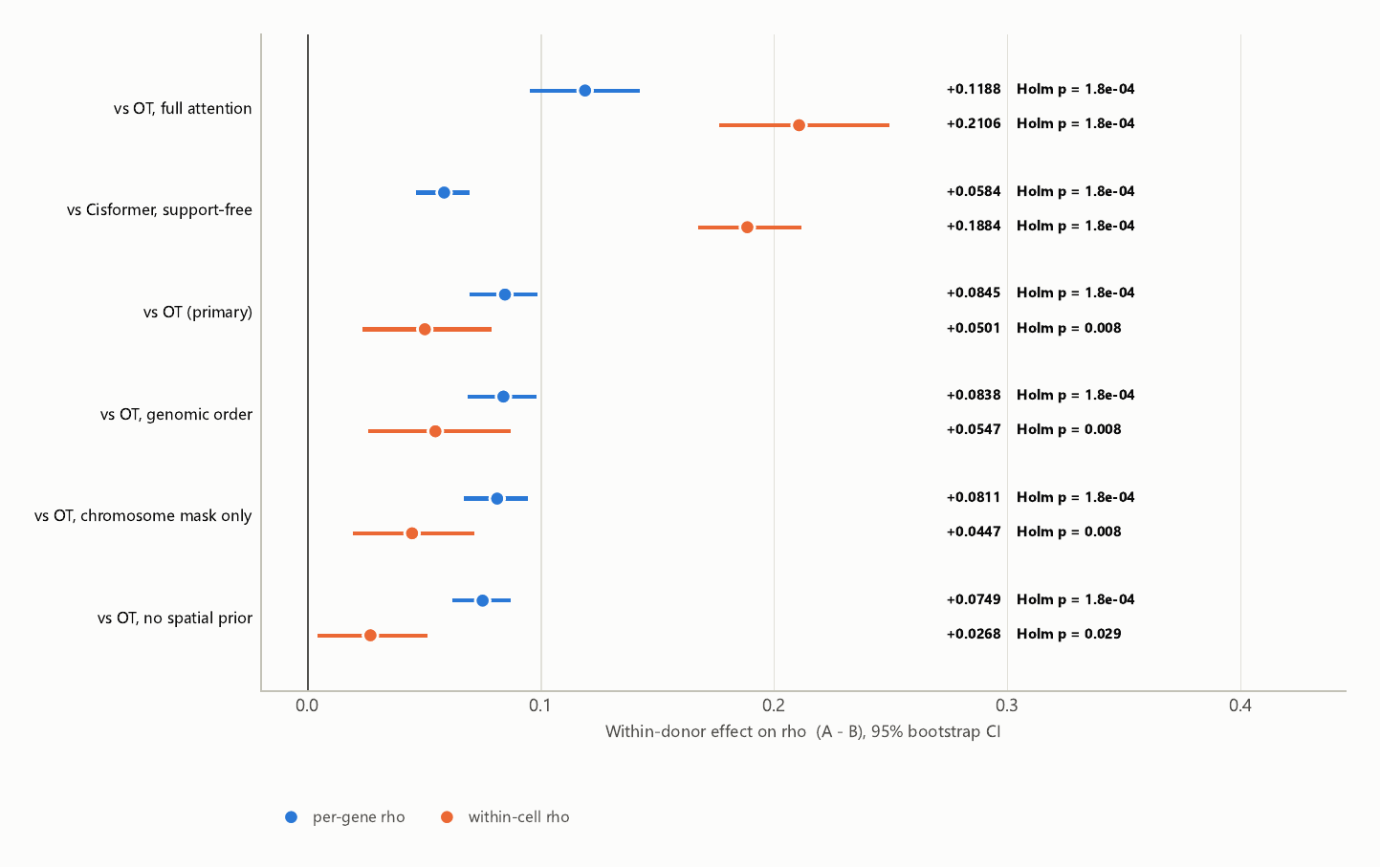}
  \caption{\textbf{Pipeline ranking.} Within-donor difference between scButterfly-B and
  each transformer pipeline (the estimand of the primary inference, not the pooled
  difference of Table~\ref{tab:benchmark}) with 95\% donor-bootstrap intervals and the
  stored Holm-adjusted $p$-values from \permDescModelRanking; Holm correction within the
  six-contrast family and within metric. All sixty paired fold comparisons favour
  scButterfly-B. \ifdefined\rankingNoteNeeded A dagger marks a percentile interval that
  excludes its own point estimate; such intervals are reported as computed together with
  the bootstrap bias and leave-one-donor-out diagnostics released with the
  results.\else Every percentile interval contains its point estimate and bootstrap biases
  are below $3 \times 10^{-4}$; in the original pooled raw-row analysis the scButterfly-B
  versus Cisformer per-gene interval excluded its estimate (bias $-0.021$), and that
  analysis is released with the inference outputs.\fi}
  \label{fig:ranking}
\end{figure}

The ranking family was declared after we saw the ablation results, and we say so. Within
it, scButterfly-B beats each of the six transformer configurations on both metrics in
every one of the sixty paired fold comparisons, with no reversal
(Figure~\ref{fig:ranking}). Figure~\ref{fig:ranking} reports
the within-donor paired differences that the primary inference tests, which are smaller
than the pooled differences in Table~\ref{tab:benchmark} because they exclude the
between-study component of the rank structure; the two must not be read as the same
quantity. The within-donor per-gene margins run from $\deltaScbVsOmicNoSpatialPriorPg$
against the best transformer pipeline to $\deltaScbVsOmicFullAttentionPg$ against the
full-attention arm, all with intervals well clear of zero. On within-cell correlation the
picture is less lopsided: the within-donor margin over the no-spatial-prior variant is
$\deltaScbVsOmicNoSpatialPriorWc$ with a bootstrap interval reaching down to
$\ciLoScbVsOmicNoSpatialPriorWc$ and $p_{\mathrm{Holm}} = \pholmScbVsOmicNoSpatialPriorWc$,
the weakest result in the analysis and the only ranking contrast whose significance
differs between the primary analysis and the pooled rank-swap sensitivity analysis
($p = 0.052$ there). scButterfly's decisive advantage is on per-gene
correlation, that is, on tracking which cells have more or less of a given gene, rather
than on the shape of the profile within a cell. Put differently, the four Perceiver
variants of \OT{} are within \gapWcPercMin--\gapWcPercMax{} of scButterfly on profile
shape, Cisformer and the full-attention arm trail by \gapWcOtherMin--\gapWcOtherMax, and
every transformer pipeline is \gapPgMin--\gapPgMax{} behind on cross-cell variation. These
are pooled performance scores; the inferential quantities are the within-donor differences.

\FloatBarrier
\subsection{The three families emit very different numbers of genes}

\begin{figure}[!htb]
  \centering
  \includegraphics[width=0.75\textwidth]{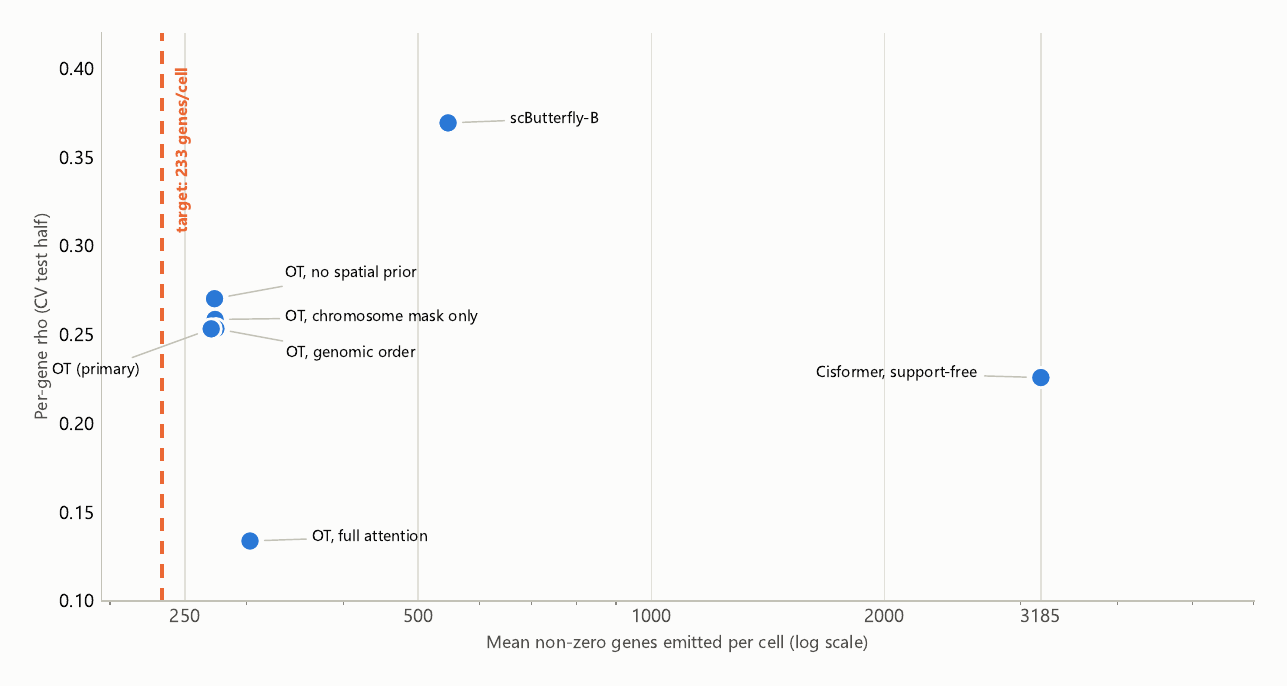}
  \caption{\textbf{Genes emitted versus accuracy.} Mean number of non-zero genes
  emitted per cell (log axis) against per-gene correlation on the cross-validation test
  half. The dashed line is the target's mean of \targetNonzero{} expressed genes per cell.
  This is a descriptive view of output sparsity, not a calibration analysis.}
  \label{fig:emission}
\end{figure}

The three model families produce very different numbers of non-zero genes per cell
(Figure~\ref{fig:emission}). The target contains \targetNonzero{} non-zero genes per cell
on average. \OT{} emits about \emitPrim, scButterfly \emitScb, and Cisformer, whose
output is a softmax expectation over abundance classes, is dense and emits all 3{,}185.
Matching the target count is not evidence of correct support or calibrated
probabilities, and we make no such claim. What the counts do show is that the per-gene
ranking is not a consequence of how many genes a model attempts: Cisformer attempts every
gene and still scores below every Perceiver variant of \OT{}. Whether scButterfly's
advantage comes from identifying which genes are expressed or from ordering cells
correctly given the support cannot be separated with these metrics; a support
precision-recall analysis on a common gene mask is the experiment that would separate
them, and we list it as future work. The coverage-conditioned per-gene correlations in the
released tables (\condScb{} for scButterfly, \condPrim{} for the primary \OT{}) are
computed on each model's own varying-gene set and are therefore not a controlled
comparison.

\FloatBarrier
\subsection{The same-atlas holdout reproduces the ordering}

Every run was also scored on the \nCellsExternal-cell holdout, which is disjoint from
both halves of every fold but drawn from the same 19 studies
(Table~\ref{tab:external}). The ordering is the same as on the cross-validation half on
within-cell correlation and the same on per-gene correlation except that the primary and
genomic-order variants of \OT{}, which differ by less than 0.001 on both sets, exchange
places. Every value is within \extMaxDiff{} of its cross-validation counterpart. This is
corroboration of the fold estimates rather than independent evidence of generalisation;
it inherits the split-design caveat above.

\begin{table}[!htb]
  \centering
  \small
  \caption{\textbf{Fixed same-atlas holdout (\nCellsExternal{} cells, disjoint from all
  folds).} Mean $\pm$ SD over the ten trained models of each configuration.}
  \label{tab:external}
  \begin{tabular}{lccc}
    \toprule
    Model & $\rhoG$ & $\rhoC$ & Coverage \\
    \midrule
    \input{generated/table_external.tex} \\
    \bottomrule
  \end{tabular}
\end{table}

\FloatBarrier
\subsection{Per-gene accuracy is low on two withheld cell types}

\begin{figure}[!htb]
  \centering
  \includegraphics[width=0.8\textwidth]{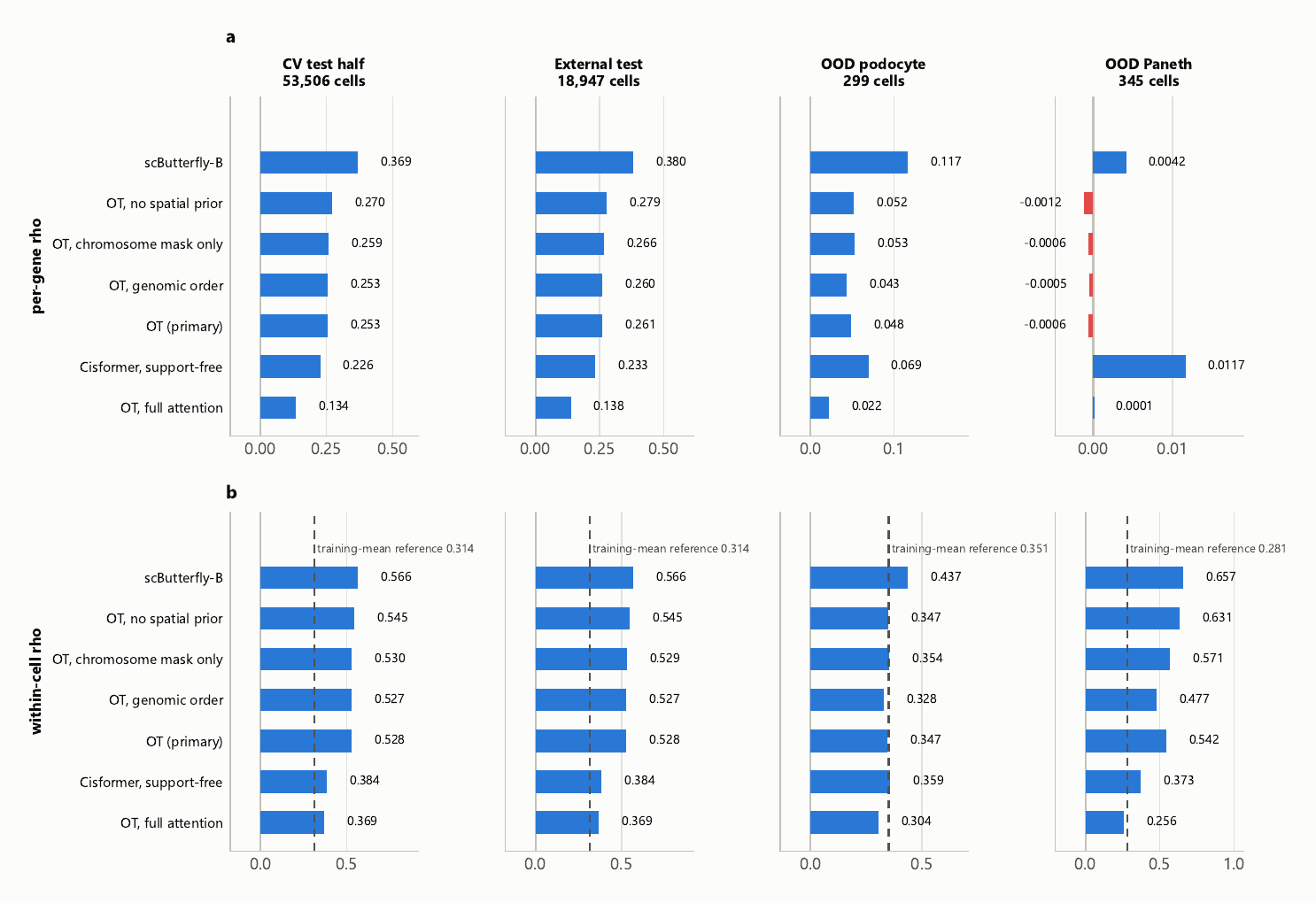}
  \caption{\textbf{Accuracy across evaluation sets.} (a) Per-gene and (b)
  within-cell correlation on the cross-validation test half, the same-atlas holdout and
  the two withheld cell types. Red bars are negative; the dashed line is the training-mean
  profile, which uses no chromatin input. Each panel has its own axis, so values should be
  compared within a panel rather than across.}
  \label{fig:generalization}
\end{figure}

Per-gene correlation is low on both withheld cell types
(Figure~\ref{fig:generalization}; Table~\ref{tab:ood}). On Paneth cells every pipeline is
at or indistinguishable from zero and \nNegPaneth{} of the seven are negative. On podocytes
the best pipeline, scButterfly-B, reaches \pgPodBest, about a third of its in-distribution
value, and the transformer pipelines reach \pgPodTfMin--\pgPodTfMax.
\ifdefined\whSeenMedianPod
Two controls computed from the saved predictions put these values in context
(Table~\ref{tab:withheld_controls}; Methods), and they separate the two withheld sets.
First, a per-gene correlation computed within one cell type is low for seen types too:
matched-size subsets ($n = \nPodocyte$) of the \whSeenTypes{} seen annotations with
enough cells in the cross-validation test half give a median within-type per-gene
correlation of \whSeenMedianPod{} for the primary \OT{} (10th to 90th percentile
\whSeenQlo--\whSeenQhi). The podocyte value lies above the median of those seen-type
subsets (\whFracSeenBelowPod{} of them score lower), so on podocytes the pipelines do
about as well as they do within a seen cell type of the same size; the drop from the
pooled in-distribution value reflects the within-type nature of the statistic more than
a failure to transfer. The Paneth value, by contrast, lies below nearly every seen-type
subset (\whFracSeenBelowPan{} score lower). Second, shuffling each pipeline's prediction
rows among the cells of a withheld set, which keeps its outputs but destroys their
assignment to cells, gives per-gene correlations of \whNullPgPod{} on podocytes and
\whNullPgPan{} on Paneth cells. On podocytes the observed value exceeds the shuffle null
in \whPodAboveNull{} of ten folds for the primary \OT{}, and in ten of ten for every
pipeline; on Paneth cells it does so in \whPanAboveNull{} of ten for the primary \OT{} and
for every other \OT{} arm, in seven of ten for scButterfly-B, and in ten of ten for
Cisformer, whose Paneth value of \pgPanCis{} is tiny but consistently above chance. So the
pipelines carry some cell-specific information about podocytes and essentially none
about Paneth cells, where the claim of an absent signal is supported by the shuffle
control.
\fi

The within-cell picture differs between the two sets and must not be pooled, and it is
where the no-input reference earns its place (Figure~\ref{fig:generalization}b). On
podocytes, \nBelowMeanPodocyte{} of the seven models, including the primary \OT{}, score
no higher than the training-mean profile's \wcPodMean. Read without that reference,
within-cell values of 0.33--0.35 would look like moderate success; with it, they are
consistent with the models reproducing shared profile structure rather than anything
podocyte-specific. On Paneth cells every model except the full-attention arm scores well
above the reference, with scButterfly at \wcPanScb; profile shape is partly recovered
there even though cell-to-cell variation is not.

Several things limit what this experiment can say. The two withheld sets are small
(\nPodocyte{} and \nPaneth{} cells), come from single studies that are themselves
represented in training, and have far fewer evaluable genes (\evaluablePod{} and
\evaluablePan{} against \evaluableCV{} in a cross-validation half), so the per-gene
statistic is not on the same footing as the in-distribution one. The ileum study is an
engineered cell line rather than primary tissue (Supplementary Table~\ref{tab:studies}).
The experiment is a cell-type holdout within seen studies, not a held-out donor or
held-out study experiment, and nothing here speaks to held-out-donor generalisation. The
claim we make is correspondingly narrow: the observed per-gene correlations in these two
withheld populations are low, and the controls above describe how low relative to seen
types of the same size and relative to a cell-agnostic shuffle of each pipeline's own
outputs.

\begin{table}[!htb]
  \centering
  \small
  \caption{\textbf{Withheld cell types.} Mean over the ten trained pipelines. Both sets
  were removed from the initial partition before the cross-validation pool was formed and
  are never seen in training or validation.}
  \label{tab:ood}
  \begin{tabular}{lcccc}
    \toprule
    & \multicolumn{2}{c}{$\rhoG$} & \multicolumn{2}{c}{$\rhoC$} \\
    \cmidrule(lr){2-3} \cmidrule(lr){4-5}
    Model & Podocyte ($n{=}\nPodocyte$) & Paneth ($n{=}\nPaneth$) & Podocyte & Paneth \\
    \midrule
    \input{generated/table_ood.tex} \\
    \bottomrule
  \end{tabular}
\end{table}

\ifdefined\whTableReady
\begin{table}[!htb]
  \centering
  \small
  \caption{\textbf{Controls for the withheld cell types, from saved predictions.} For
  each pipeline: the observed per-gene correlation on the withheld set (mean over ten
  folds); the mean of the same metric after shuffling the pipeline's prediction rows among
  the cells of that set (200 shuffles per fold); the number of folds in which the observed
  value exceeds the 97.5th percentile of its shuffle null; and the median per-gene
  correlation obtained on seen cell types from the cross-validation test half when
  subsets of the same size are scored within one annotation (five subsets per annotation
  with at least 345 cells).}
  \label{tab:withheld_controls}
  \setlength{\tabcolsep}{2.5pt}\scriptsize
  \begin{tabular}{lcccc|cccc}
    \toprule
    & \multicolumn{4}{c|}{Podocyte ($n{=}\nPodocyte$)} & \multicolumn{4}{c}{Paneth ($n{=}\nPaneth$)} \\
    Pipeline & obs. & shuffled & folds $>$ null & seen median & obs. & shuffled & folds $>$ null & seen median \\
    \midrule
    \input{generated/table_withheld_controls.tex} \\
    \bottomrule
  \end{tabular}
\end{table}
\fi

\FloatBarrier
\section{Discussion}

The most useful result here is a negative one about design intuition. The two genomic
inductive biases built into \OT{} were motivated by the biology: regulatory elements act
mostly in \emph{cis}, and proximity to a gene is the single most predictive feature for
enhancer-gene linkage. Encoding that as a chromosome mask and a distance penalty on
latent-to-peak attention nonetheless made the model worse, reproducibly, on both metrics.
We think the explanation is that the prior is in the wrong place. The latents are not
genes; they are a learned compression of the whole accessibility profile, and the decoder
that emits genes never sees genomic coordinates at all. A latent anchored to a stretch of
chromosome 7 is forbidden from integrating a cell-state signal that happens to be spread
over chromosome 12, even though that signal is what the decoder needs. A cis prior would
have to be expressed between peaks and the specific genes being emitted, which is a
different architecture, and nothing here speaks against biological priors implemented
that way. The null result for output order points the same way: the decoder does not
appear to exploit genomic adjacency between consecutive emissions, so ordering them
genomically buys nothing detectable.

The one encoder change that helped, replacing full self-attention over 8{,}192 sparse
tokens with a Perceiver, is also the least biologically specific. Its benefit is partly
optimisation: the full-attention arm is unstable across folds and stops early, and its
within-cell correlation is barely above the no-input reference. Because the contrast
also changes the decoder memory and the parameter count, and because one shared training
recipe does not guarantee that both encoders are comparably tuned, we read it as
evidence that the implemented Perceiver configuration trains better, not as a general
statement about compression.

The benchmark result is equally clear and more limited in what it licenses. The
scButterfly-B pipeline beats every transformer pipeline we trained, in every fold, on
both metrics, at about a quarter of the training time. We do not conclude that
autoencoders beat transformers for this task, for three reasons. First, it is one
autoencoder-based pipeline against one transformer family plus one adapted external
transformer; the class-level comparison has $n = 1$ on each side. Second, the contrast is
not controlled: scButterfly brings its own preprocessing, TF-IDF transformation, panel-
normalised training target, output space and training objective, so nothing in the
comparison isolates ``autoencoder'', and the three pipelines are not trained on
equivalent targets. Third, the three families produce outputs of very different
sparsity, and our metrics cannot separate support identification from abundance
ordering given the support. What the data support is narrower: on this atlas, under
support-free prediction with a shared feature vocabulary and a common scoring target,
the scButterfly-B pipeline as implemented is the strongest tested, and its advantage lies
in cross-cell per-gene variation rather than within-cell profile shape, where the best
transformer pipeline is within \gapWcPercMin.

The withheld-type controls also change how the two withheld sets should be read. A
per-gene correlation computed within one cell type has far less cell-to-cell variation
to explain than the pooled multi-tissue statistic, and the matched-size seen-type subsets
show that values of a few hundredths are the norm, not the exception, for every pipeline.
By that yardstick podocytes are unremarkable: the pipelines recover about as much
cell-specific signal there as within a seen type of the same size, and more than a
shuffle of their own outputs. Paneth cells are the genuine failure, below nearly every
seen-type subset and at the shuffle null for every \OT{} arm; whether that reflects the
engineered origin of the ileum sample, the distinctive secretory programme, or a true
limit of chromatin-based prediction cannot be decided from one population.

The reference predictors deserve more attention than they usually get. A constant
written on the true support scores \pgOracle; a dense model evaluated that way cannot be
distinguished from a lookup of which genes are on. Our support-free Cisformer
reformulation, which queries all 3{,}185 genes with a fully open decoder mask and
supervises the zero class, is stricter than the published protocol, and its placement
near the bottom of the ranking should be read as a statement about that stricter task,
not about the published model. The training-mean profile is just as important in the
other direction: \nBelowMeanPodocyte{} of seven models on podocytes achieve within-cell
correlations that look respectable in isolation and are in fact no higher than what a
model with no input achieves.

Several limitations shape how far these results travel. Feature selection used the whole
atlas, including the RNA values of cells later held out, and scButterfly's own
preprocessing statistics were fitted on the pooled object, so the protocol is
transductive with respect to the feature space and the baseline's normalisation; a
vocabulary fixed across folds is not an independently defined vocabulary, and we make no
claim of fully inductive evaluation. The three pipelines are trained on different native
representations of the same counts, and rank metrics do not make those equivalent for the
per-gene statistic (Methods); the ranking is a ranking of pipelines on a common scoring
target, not of architectures on equivalent targets. The split is at cell level, so cells
from one donor appear in both halves of every fold; the cross-validation numbers are
in-distribution estimates and say nothing about held-out-donor generalisation. The
withheld cell types are the only evidence about transfer, and that evidence is limited in
the ways described above. The decoding cap of 384 steps was reached in a substantial
fraction of cells; re-decoding at 768 steps raised the \OT{} per-gene scores by about
\capDeltaPgTest{} on a test subset, which does not change any ranking but means the
reported \OT{} values are slight underestimates of what the checkpoints can produce. The ranking family was declared post hoc. Each model was
early-stopped on its own validation loss, so the selection rule is harmonised but the
quantity is not. And the whole study is one atlas with one feature vocabulary; whether
the ordering holds under a different peak set or a different normalisation is an open
question.

We think the practical recommendations are simple. Benchmarks of chromatin-to-expression
translation should state whether target support is supplied at inference, include a
no-input profile reference and a support-only diagnostic, report both per-gene and
within-cell correlation, charge zero for genes a model does not vary on or report
coverage alongside, recover and use the true independent sampling unit in the inference,
and prespecify architectural contrasts before looking at results. None of these is hard,
and in our hands each one changed a conclusion we would otherwise have drawn.

\section{Methods}

\FloatBarrier
\subsection{Paired multiome atlas}

The data are a paired snATAC-seq/snRNA-seq compendium of human tissues assembled as a
Seurat weighted-nearest-neighbour object~\citep{hao2021seurat,stuart2021signac} and
distributed with the EpiFoundation resource~\citep{wu2025epifoundation}; the underlying
libraries come from ENCODE~\citep{encode2020} and GEO. The object contains 133{,}549
nuclei, 36{,}601 measured genes and 57 cell-type annotations from 19 studies, each study
contributing one tissue (Supplementary Table~\ref{tab:studies}). The RNA matrix is stored
in Seurat LogNormalize form, $\log(1 + c_{ig}\,/\,N_i \times 10^4)$ for count $c_{ig}$
and library size $N_i$; we verified by exact inversion on the stored matrix that integer
counts are recovered (maximum deviation $6\times 10^{-12}$), and the released
preparation route repeats that check before writing the library-size file. The
accession identifier is the only grouping variable in the object and is one-to-one with
tissue. Donor identity was recovered after the fact from the ENCODE experiment records
(biosample donor accession) and the GEO sample records; the mapping is released with the
code. The 11 ENCODE accessions correspond to 8 donors; the 8 GEO studies are single
samples without cross-series donor identifiers and are treated as one unit each, giving
16 independent units.

We applied the following cell-level filters: ATAC fragment count $\ge 1{,}000$, ATAC
features $\ge 500$, nucleosome signal $\le 4$, TSS enrichment $\ge 2$, RNA UMI count
$\ge 500$, RNA features $\ge 200$, and mitochondrial fraction $\le 20\%$. The library-size
range in the object (1{,}001--24{,}997 UMIs) indicates that upstream filtering had
already been applied; our filters removed 154 cells on TSS enrichment and 6{,}791 on
mitochondrial fraction, leaving 126{,}604.

\FloatBarrier
\subsection{Shared feature spaces}

The peak vocabulary is the 50{,}000 most broadly accessible peaks by number of cells with
a non-zero count, computed over all 133{,}549 nuclei, sorted by chromosome and start
coordinate, covering chromosomes 1--22, X and Y. Accessibility is binary: every non-zero
entry of the stored ATAC matrix is 1. Cells have on average 4{,}262 accessible
vocabulary peaks (median 3{,}657).

The gene vocabulary was constructed by selecting 5{,}000 highly variable genes with
scanpy~\citep{wolf2018scanpy} on the whole RNA object and intersecting them with
protein-coding and lncRNA genes from GENCODE v44 basic~\citep{frankish2021gencode} that
have unique symbols and coordinates, which leaves 3{,}185 genes ordered by chromosome and
transcription start site. Both vocabularies were therefore fixed on the complete atlas
before any partition. The selection is transductive: the highly-variable-gene statistics
were computed from the RNA values of every cell, including the cells of the fixed holdout,
the withheld cell types and every future test half, so held-out RNA informed the choice of
target genes; the peak vocabulary likewise used the accessibility of every cell. The
vocabulary is identical for every pipeline and fold, which removes it as a source of
between-pipeline differences, but a vocabulary fixed across folds is not an independently
defined vocabulary, and no part of the evaluation is fully inductive. In this space the target matrix is 92.7\% zero and a cell
expresses on average 231 vocabulary genes (\targetNonzero{} on the cross-validation test
halves).

\FloatBarrier
\subsection{Target and native input representations}

The scoring target for every model and every metric is the stored LogNormalize value
restricted to the 3{,}185 vocabulary genes; no further transformation is applied at
scoring time. The three models are trained on different representations of the same
underlying counts, and we state them explicitly because Spearman correlation across
cells is invariant only to a common monotone transformation of a gene's values, not to a
transformation whose scale differs from cell to cell. \OT{} regresses the stored
LogNormalize value directly. scButterfly receives integer counts recovered by inverting
the LogNormalize transform with each cell's library size, restricted to the 3{,}185-gene
panel, and applies its own total-count normalisation over that panel followed by
$\log(1+x)$; the per-cell scale factor is therefore the panel total rather than the full
library size. Cisformer receives $\log(1 + \text{count})$ discretised into eight classes
relative to each cell's own maximum. Rank metrics computed against the common target
compare each model's native output to one reference, but a per-cell rescaling can
reorder cells for a given gene, so the per-gene ranking is a comparison of methods as
run on their native representations rather than of architectures on an identical
target. A common-target sensitivity analysis is the experiment that would remove this
ambiguity and is listed as future work.

\FloatBarrier
\subsection{Held-out sets and cross-validation design}

An initial stratified random split by annotation (70/15/15, seed 42) was made once. All
podocytes ($n = \nPodocyte$) and Paneth cells ($n = \nPaneth$) were then removed from
every partition of that split and stored as two withheld sets. The \nCellsExternal{}
cells remaining in the 15\% test partition form the fixed same-atlas holdout. The union
of the remaining training and validation partitions, 107{,}013 cells from 19 tissues and
55 annotations, is the cross-validation pool.

For each of five repetitions (outer seeds 100, 200, 300, 400, 500) the pool was dealt
into two halves of 53{,}506 and 53{,}507 cells: within each tissue $\times$ annotation
stratum the cells were permuted and alternated between halves, with the larger side of
odd-sized strata alternating across strata so that the halves balance. Direction 0 trains
on half A and tests on half B; direction 1 reverses them. Within the training half, an
inner validation set of 10\% per stratum (5{,}346--5{,}364 cells; seed = outer seed +
direction) was carved out and used only for early stopping; the remaining
48{,}143--48{,}160 cells were used for gradient updates. The generator asserts that
train, validation and test are pairwise disjoint and that the test half is disjoint from
the holdout and withheld sets. Each split file carries a SHA-256 of its membership in
its metadata, and the manifest of those hashes is released.

\FloatBarrier
\subsection{\OT{} architecture}

\paragraph{Input tokens.} Each accessible peak in a cell is one token. A token carries
the peak's vocabulary index, its chromosome (1--22, X, Y) and the integer midpoint of
the peak in base pairs. The input embedding is the sum of a learned peak embedding
($50{,}001 \times 256$, index 0 reserved for padding), a learned chromosome embedding
($32 \times 256$) and a linear embedding of the accessibility value, which is constant
because accessibility is binary. No positional embedding is added; coordinates enter
only through the attention bias described below. Sequences are capped at 8{,}192 tokens.
Cells with more accessible peaks (8.4\% of cells) contribute a uniform random subset of
8{,}192 without replacement, drawn from a seeded hash; at this cap 94.9\% of
accessibility events are retained. The intended protocol redraws the training subset
every epoch from a hash of (model seed, epoch, cell name). In the runs reported here the
data-loader workers were persistent and did not receive the epoch update, so each run
used one fixed subset per above-cap training cell, drawn with the epoch-1 seed; the
released code recreates workers each epoch and a regression test covers it. At evaluation
the subset is fixed from a hash of (split hash, cell name), identical across the five
\OT{} arms and across repeated evaluations of one checkpoint.

\paragraph{Perceiver encoder.} The encoder holds 256 learned latent vectors of width 256.
Two blocks of cross-attention (4 heads, 64 dimensions per head, pre-LayerNorm) from the
latents to the peak tokens, each followed by a feed-forward layer (256--1024--256, GELU,
dropout 0.1), are followed by two blocks of latent self-attention with the same
feed-forward layer~\citep{jaegle2021perceiver,vaswani2017attention}.

\paragraph{Genomic attention prior.} Each latent is assigned a fixed genomic anchor
before training: at least one latent per chromosome, the remainder allocated to
chromosomes in proportion to their peak counts by largest remainder, and within a
chromosome the $k$ anchors placed at the $(i + \tfrac12)/k$ quantiles of that
chromosome's peak midpoints, snapped to an observed peak. Anchors are buffers and are not
trained. In the primary configuration the cross-attention logits from latent $m$ to peak
$l$ are set to $-\infty$ if the two are on different chromosomes and otherwise reduced by
$s_h \,|a_m - p_l| / 10^6$, where $a_m$ and $p_l$ are the anchor and peak midpoints in
base pairs and $s_h$ is a fixed per-head slope of $1/2^{h+1}$ for heads $h = 0..3$, that
is 0.5, 0.25, 0.125 and 0.0625 per megabase, following the geometric slope schedule of
ALiBi~\citep{press2022alibi}. A latent whose chromosome has no accessible peak in a cell
receives no cross-attention update and passes through the residual unchanged. The
\emph{chromosome mask only} variant keeps the cross-chromosome block and drops the
distance term; the \emph{no spatial prior} variant masks only padding and is invariant to
coordinates. The prior adds no parameters.

\paragraph{Full-attention variant.} The \emph{full attention} arm replaces the Perceiver
with a two-layer standard transformer encoder (4 heads, feed-forward 1024, GELU,
pre-norm) applying self-attention over all 8{,}192 peak tokens with a padding mask and no
coordinate bias, and the decoder cross-attends to all peak positions rather than to 256
latents. Attention is computed with PyTorch scaled-dot-product kernels so the
$8{,}192^2$ matrix is not materialised. Because model construction consumes random
draws differently in the two encoders, a shared seed does not make the decoder
initialisation identical between this arm and the Perceiver arms.

\paragraph{Autoregressive decoder.} A two-layer causal transformer decoder (4 heads,
feed-forward 1024, GELU, pre-norm) attends to the encoder memory. Its vocabulary has
3{,}188 tokens: padding, beginning-of-sequence, end-of-sequence and the 3{,}185 genes.
Three linear heads on each position predict the next gene token, a mean $\mu$ and a
log-variance for the abundance of the gene at that position. Expression is continuous;
there is no binning. The decoder has no positional embedding; order is conveyed by the
causal mask and the previously emitted tokens. The training target for a cell is its
non-zero vocabulary genes in a fixed order, truncated to 511 genes (affecting about 1\%
of cells), followed by end-of-sequence. In the primary configuration the order is a
single random permutation of the 3{,}185 genes (seed 1234), identical for every cell,
fold and arm; the \emph{genomic order} variant uses chromosome-then-TSS order instead.
The Perceiver configurations have 19.78 million parameters, 12.8 million of them in the
peak embedding; the full-attention configuration has 18.13 million.

\paragraph{Objective.} With teacher forcing, the loss is
$0.3\,\mathcal{L}_{\mathrm{CE}} + \mathcal{L}_{\mathrm{NLL}}$, where
$\mathcal{L}_{\mathrm{CE}}$ is token-averaged cross-entropy over the gene vocabulary
(padding ignored, end-of-sequence supervised) and $\mathcal{L}_{\mathrm{NLL}}$ is the
Gaussian negative log-likelihood $\tfrac12\,[\log\sigma^2 + (y-\mu)^2/\sigma^2]$ of the
LogNormalize target at gene positions, with $\log\sigma^2$ clamped to $[-10, 10]$. The
same teacher-forced loss on the inner validation set is the early-stopping criterion.

\FloatBarrier
\subsection{Training}

All five \OT{} arms share one training recipe: AdamW~\citep{loshchilov2019adamw} with
learning rate $3\times10^{-4}$ and weight decay 0.01; a per-step schedule with linear
warm-up over the first 3\% of the nominal 100-epoch budget followed by cosine decay;
batch size 32 without gradient accumulation; mixed precision with loss scaling; gradient
norm clipping at 1.0; dropout 0.1 throughout. Validation loss is evaluated after every
epoch on the complete inner validation set, and training stops after five epochs without
improvement; the best checkpoint is reloaded before any held-out decoding. Runs stopped
after 13--29 epochs. The model seed for fold $(r, d)$ is $7000 + 10r + d$ and is shared
by the five \OT{} arms and by Cisformer; cuDNN is set to deterministic mode and
data-loader workers are seeded. Free-running validation decoding and the two
correlation metrics were logged every epoch for the record but were not used for
selection.

Every run used one NVIDIA L40S (48~GB) under Slurm with a 36-hour limit. Peak reserved
GPU memory in a separate synthetic probe at batch 32 and 8{,}192 tokens was 9.7~GB for
the two configurations with the distance prior, 8.9~GB for the two without, and 8.6~GB
for full attention; these are probe measurements, not peaks of the training runs. The
\OT{} training timer covers the epoch loop including the teacher-forced validation loss
and the free-running validation decode performed every epoch; it excludes data loading
set-up and held-out inference. Mean training wall-clock per fold was \trainHPrim~h for
the primary configuration (23.5 epochs on average, about 500~s per epoch) and
\trainHFull~h for full attention (16 epochs at about 1{,}440~s per epoch). Free-running
decoding of a \nCellsTestHalf-cell test half took about 2{,}500~s for Perceiver arms and
9{,}400~s for full attention.

\FloatBarrier
\subsection{Inference}

Decoding is support-free: the target RNA is dropped from the batch before the model is
called. The encoder is run once per cell and the decoder generates greedily from the
beginning-of-sequence token, taking the arg-max gene at each step with no sampling, beam
search, temperature or repetition mask, until every cell in the batch has emitted
end-of-sequence or 384 steps have elapsed. The prediction matrix is initialised to zero;
for each emitted gene the predicted abundance is $\max(0, \mu)$ at that step, and the
first emission of a gene is kept if the same gene is emitted again. The implementation
that produced the stored matrices tested whether a gene had been seen by whether its
current prediction was zero, so a first emission clipped to zero could be overwritten by
a later duplicate; the released code keeps an explicit seen-mask, and the stored decoder
streams (tokens, means, log-variances, sequence lengths and end-of-sequence positions)
allow the matrices to be rebuilt under the corrected rule without a GPU. \ifdefined\nativeCellsChanged
Rebuilding all \nativeRunsChecked{} cross-validation test-half matrices this way changed
\nativeEntriesChanged{} entries in \nativeCellsChanged{} cells; the 384-step cap was
reached before end-of-sequence in \nativeCapHitPct\% of cells and cells emitted
\nativeDupsPerCell{} duplicate tokens on average.
\else
The number of cells affected, the frequency with which the 384-step cap was reached before
end-of-sequence, and the number of duplicate emissions are reported with the released
results.
\fi
\ifdefined\capRuns
Because the cap was reached in a large fraction of cells, we re-decoded every checkpoint
(\capRuns{} runs) at a predetermined longer cap of 768 steps on both withheld sets and a
seeded 5{,}000-cell subset of each test half. Lengthening the cap changed per-gene
correlation by \capDeltaPgPod{} on podocytes (cap reached in \capHitShortPod\% of cells at
384 steps, \capHitLongPod\% at 768), by \capDeltaPgPan{} on Paneth cells and by
\capDeltaPgTest{} on the test subset (mean over runs; largest absolute change on
podocytes \capDeltaPgMaxPod).
\fi A determinism check confirmed that decoding the same checkpoint twice at batch
4 gives bitwise-identical predictions and that batch 7 with two workers gives identical
token streams and means within $10^{-4}$.

\FloatBarrier
\subsection{scButterfly-B baseline}

scButterfly~\citep{cao2024scbutterfly} version 0.0.9 was run unmodified in its B
configuration, that is, without the cell-type or cluster-based data augmentation of the
T and C variants. Because the package expects raw counts, we recovered integer counts
from the LogNormalize matrix as $\mathrm{round}(\mathrm{expm1}(x)\,N_i / 10^4)$ and
supplied all 50{,}000 vocabulary peaks of the binary ATAC matrix in chromosome order,
with the per-chromosome peak counts required by its chromosome-split ATAC encoder. Its
own preprocessing was then applied to one object containing the training, validation and
all held-out cells together: total-count normalisation over the 3{,}185-gene panel and
$\log(1+x)$ for RNA with no highly-variable-gene selection, and binarisation,
low-frequency peak filtering (\texttt{fpeaks} = 0.005), TF-IDF and normalisation for
ATAC. The peak filter, inverse document frequencies and normalisation constants are
therefore fitted on the pooled object including held-out cells; no held-out RNA target
enters any gradient update, but this preprocessing is transductive. Training used the
package defaults for architecture and optimisation (dual VAEs with a chromosome-split
ATAC encoder, 128-dimensional translator, two discriminators, Adam at $10^{-3}$ for the
autoencoders and SGD at $5\times10^{-3}$ for the discriminators, MSE/BCE reconstruction
and KL terms with the package's loss weights) with batch size 64, 30 epochs of RNA and
ATAC pre-training, up to 60 translator epochs, and early stopping on its own validation
reconstruction loss with patience 5. The package resets all random generators to its
internal seed 19193 in its constructor and training call, so the fold seed recorded in
the run manifests was not the effective seed for scButterfly; the ten runs differ in
their data, not in their initialisation seed. The released wrapper now re-seeds after
construction and passes the fold seed explicitly. The job logs show that the
integrative stage early-stopped in all ten runs, so the package's best-validation state
was restored in every reported run. The inner validation cells were placed at the head
of the package's test list so they were used only for checkpoint selection and never for
gradient updates. Predictions for the test, holdout and withheld cells were taken from
the ATAC-to-RNA path, which returns an $n \times 3{,}185$ matrix in vocabulary order,
and written to the shared prediction store. scButterfly's output lives in its own
normalised space, so mean-squared error is not reported for it. The scButterfly timer
covers preprocessing and all three training stages and stops before the final decoded
validation pass; the job required 256~GB of host memory for the TF-IDF step and the mean
training time was \trainHScb~h per fold on the same L40S.

\FloatBarrier
\subsection{Cisformer, support-free}

Cisformer~\citep{ji2025cisformer} tokenises accessible peaks and decodes per-gene
abundance classes with a transformer decoder. In the published ATAC-to-RNA protocol the
decoder is queried only at the genes that are non-zero in the target cell, and the zero
class is excluded from the loss. We kept the released architecture (model width 280,
four decoder layers with self-attention, cross-attention and feed-forward sub-layers,
7 heads of dimension 140, feed-forward multiplier 4, dropout 0.1; a seven-digit decimal
embedding of the peak index in place of an encoder; eight abundance classes, where class
0 is zero and classes 1--7 are septiles of the cell's own maximum $\log(1 + \text{count})$)
and changed the task. The decoder is queried on all 3{,}185 genes with an all-true
decoder mask, which is asserted at inference, and the zero class is supervised with the
same class-weighted cross-entropy as the other classes (inverse-frequency weights
estimated from the first 200 training batches, zero-class weight doubled). A regression
test asserts that no target information reaches the decoder mask. Peaks beyond the
8{,}192 cap are subsampled deterministically from a seeded hash of the cell and split;
in the reported runs that hash used a Cisformer-specific namespace, so above-cap cells
(8.4\%) were evaluated on a different subset of their peaks than in \OT{}; the released
code uses the shared implementation and a test asserts identical subsets. Training used
Adam with AMSGrad at $5\times10^{-4}$, a step schedule multiplying the rate by 0.6 every
four epochs, mixed precision, batch size 8, at most 30 epochs, and early stopping on
validation cross-entropy with patience 5 after three warm-up epochs. The checkpoint with
the lowest validation loss was retained: the job logs of all ten runs record a checkpoint
save at each improvement in validation loss, and the saved checkpoints carry the
validation loss as their selection score and the epoch of minimum validation loss
(Supplementary Table~\ref{tab:provenance}). The point prediction for each gene is the
softmax expectation over the eight class indices; the arg-max class is stored as a native
side output. Because every gene receives a value, coverage is 1 by construction. The
Cisformer timer covers the epoch loop including the per-epoch validation loss and
validation decode; the mean training time was \trainHCis~h per fold.

\FloatBarrier
\subsection{Reference predictors}

Two reference predictors are evaluated on every fold through the same prediction-store
and scoring path as the trained models. The \emph{training-mean profile} is the per-gene
mean of the LogNormalize matrix over the fold's full training half (training plus inner
validation cells), broadcast to every test cell; it uses no test information and no
chromatin. Its per-gene correlation is zero and its coverage is zero because every column
is constant. The \emph{constant oracle-support diagnostic} writes 1 at every position
where the target is non-zero and 0 elsewhere; it uses target support and nothing else and
is the appropriate reference for any protocol that conditions on expressed-gene
identity. Oracle-support Cisformer results were run for diagnosis only and do not enter
any ranking.

\FloatBarrier
\subsection{Metrics}

One implementation scores every model. Ranks are tie-corrected average ranks and Spearman
correlation is the Pearson correlation of ranks. A vector is treated as varying if and
only if its maximum differs from its minimum, evaluated on the stored values; the
original implementation used a standard-deviation threshold that float32 accumulation
could exceed for genuinely constant columns, and every reported number was recomputed
after the correction. For \emph{per-gene} correlation, a gene is evaluable if its target
varies across the scored cells; the correlation is computed across cells for every
evaluable gene whose prediction also varies, and set to 0 for evaluable genes whose
prediction is constant. The reported value is the mean over evaluable genes
(\evaluableCV{} of 3{,}185 on the cross-validation halves, \evaluableExt{} on the
holdout, \evaluablePod{} on podocytes and \evaluablePan{} on Paneth cells).
\emph{Coverage} is the fraction of evaluable genes whose prediction varies, and the
\emph{coverage-conditioned} per-gene correlation is the mean over that subset only. For
\emph{within-cell} correlation, every cell whose target and prediction both vary across
the 3{,}185 genes is scored across all 3{,}185 genes, emitted or not, and the mean over
those cells is reported; all \nCellsTestHalf{} cells of a test half are scored, with no
subsampling. The mean number of non-zero predicted and target genes per cell are also
recorded. Mean-squared error is reported only for models whose output is on the target
scale.

\FloatBarrier
\subsection{Prediction stores, provenance and verification}

Every run writes its dense prediction for each evaluation set to an HDF5 store (float32,
gzip) with the cell identifiers, study, tissue and annotation of every row, the gene
axis, native side outputs, and a manifest recording the command line, a SHA-256 of the
source tree that executed, software environment, GPU, Slurm job, model seed, checkpoint
SHA-256 and wall-clock times. Stores are written atomically and re-opened for validation
before being accepted. The canonical target for each split and evaluation set is written
once and verified to be byte-equal to the corresponding rows of the RNA matrix. A
separate scoring script re-opens every store, joins it to the canonical target on cell
identifier and recomputes all metrics; disagreements with the value computed in-process
are recorded with their cause. The 400 stores total 24.5~GB and are checksummed.

The cluster copy of the code is an rsync of the development tree without version
control, so manifests record a content hash of the executed source rather than a commit.
Two source-tree hashes appear across the 70 runs; neither corresponds exactly to a
commit, because the tree was edited on the cluster during the campaign. For the
Cisformer runs the executed selection rule was verified from the job logs and checkpoint
metadata as described above; the Cisformer runner in the released repository had, at the
time of the audit, a selection branch that could not have produced those runs, and it
has been corrected and tested. The pre-audit cluster source tree is archived alongside
the results.

\FloatBarrier
\subsection{Statistical inference}

For each prespecified contrast and metric, the fold-level difference is
$\mathrm{metric}(\mathrm{A}) - \mathrm{metric}(\mathrm{B})$ on the same test half; the
repetition-level difference is the mean of the two directions; the test statistic is the
mean of the five repetition-level differences. Because the two directions of a repetition
train on complementary halves and test on each other's training data, and the five
repetitions overlap randomly, we do not apply a $t$-test across folds.

Inference treats the donor as the independent unit (16 units; Supplementary
Table~\ref{tab:studies}). The analysis reported in the tables was revised after the
independent audit, which showed that the original procedure, exchanging the two models'
raw prediction rows within randomly chosen studies and recomputing the pooled statistic,
depends on each model's output scale: the pooled across-cell ranks of a hybrid matrix
change under a per-model rescaling even though the observed Spearman statistic does not.
The primary inference is now an exact sign-flip test on within-donor paired differences.
For each contrast, metric and donor, the metric is computed on that donor's cells alone in
every fold, the paired difference $\mathrm{A} - \mathrm{B}$ is averaged over the two
directions and the five repetitions, and the test statistic is the mean of the 16
donor-level differences. Under the null hypothesis that neither pipeline is better, each
donor's difference is equally likely to carry either sign, so the null distribution is
obtained by enumerating all $2^{16}$ sign patterns exactly; the two-sided $p$-value is the
fraction of patterns whose statistic is at least as extreme as the observed one, and it
has no Monte Carlo floor. The assumptions are that donors are independent, that the sign
of each donor's difference is exchangeable under the null, and that the within-donor
difference is the quantity of interest. The estimand differs from the pooled difference
of performance scores: between-study rank structure is excluded, so within-donor effects
are smaller in magnitude, and the two are reported separately throughout. Donors who
contributed several tissues (one contributes colon, adrenal gland and fallopian tube,
another muscle and bile duct) are single units whose cells are pooled; this respects
their dependence but does not remove the between-study structure within such a unit, so
their donor-level differences are averages over tissues. Confidence intervals are
percentile intervals from 2{,}000 bootstrap resamples of the 16 donor-level
differences~\citep{davison1997bootstrap}; the released output records the bootstrap
bias and leave-one-donor-out estimates for every contrast.

As a sensitivity analysis the pooled statistic is also tested with a cluster permutation
that exchanges per-gene rank-transformed prediction rows within donors, which is
invariant to per-model monotone rescaling and leaves the observed pooled statistic
unchanged (1{,}000 permutations and 1{,}000 cluster bootstraps per contrast); its
results, and the original raw-row analysis kept for reproducibility, are released with the
inference outputs. Monte Carlo $p$-values there are reported as
$(\#\{|T^\ast| \ge |T|\} + 1)/(n + 1)$ together with an exact binomial upper bound on the
tail probability. Holm's step-down
correction~\citep{holm1979} is applied within each family and within each metric, never
across. The \emph{ablations} family contains the four contrasts of
Table~\ref{tab:ablations} and was fixed in the analysis plan committed to the repository
on 5 August 2026 before any run of the matrix was launched; the \emph{model ranking}
family, scButterfly-B against each of the six transformer configurations, was added on
10 August 2026 after the ablation results were seen, and we call it post hoc throughout.
No external registration was filed; the commit history is the record of prespecification.
The plan also forbids attributing the primary versus full-attention difference to the
encoder, since that pair differs in the spatial prior as well.

\FloatBarrier
\subsection{Software, development process and reproducibility}

Models were implemented in PyTorch 2.5.1 with CUDA 12.1~\citep{paszke2019pytorch} under
Python 3.9 on the cluster; the exact package versions of the environment that produced
the results are recorded in the repository. Split generation, target writing, training,
scoring, aggregation, permutation inference and figure generation are separate scripts
that communicate only through the split files, prediction stores and their manifests.
Every table and every numeric macro in this paper is generated from the aggregated
metric table and the inference outputs by one script, and the prose claims that are not
themselves generated (sign counts, rank orders, ranges) are pinned by assertions in the
same script.

The code and analyses were developed with an AI coding agent under the audited vibe
coding approach described previously~\citep{meyer2026auditedvibe}: the agent wrote and
ran the pipeline under human direction, every quantitative claim was traced back to a
stored artefact, and the complete repository and manuscript were then subjected to an
independent pre-submission audit whose findings, with the resulting code changes,
regression tests, reanalyses and claim revisions, are recorded item by item in the
repository.

\section*{Data and code availability}

Code for preprocessing, split generation, all seven pipelines, the two reference
predictors, the shared metric implementation, the inference procedures, and figure
generation is at \url{https://github.com/xomicsdatascience/omic-translator-full} under
the MIT licence for original code, with upstream licences retained. The repository
includes the split manifest with a SHA-256 of every fold's membership, the study-to-donor
map, the aggregated metric table for all runs and evaluation sets, the per-fold paired
differences, the inference outputs with their raw draws, the native-stream
reconstruction, decoding-cap and withheld-type control reports, the quality control
report, the recorded cluster environment, the audit and its item-by-item response. The
input atlas is the MiniAtlas release of the EpiFoundation
resource~\citep{wu2025epifoundation} (Hugging Face dataset UCSC-VLAA/MiniAtlas, revision
275941ba); the two input files used here hash exactly to that revision, and the
preparation route from those files through quality control, feature selection,
normalisation handling and splits is scripted and verified. A compact, checksummed
release built by the repository (versioned code and environment, the ten exact split
files, gene and peak axes, the donor map, per-fold metrics for every run, unit-level
paired differences with a script that recomputes the sign-flip tests from them, figure
and table source data, run manifests, training histories, checkpoint-selection evidence
and Slurm logs, and a small prediction-and-target example that exercises the scoring
step) is deposited on figshare at \url{https://doi.org/10.6084/m9.figshare.34617930}
(release tag \texttt{v3.1-audit-2026-10-16}). The complete cell-level prediction stores, canonical
targets and checkpoints are kept in a durable internal archive whose file-level
checksums are included in the release, and are available to reviewers on request; the
release states which facts are hashed exactly and which are inferred from logs and
checkpoint metadata.

\section*{AI use disclosure}

The code, the analyses and the writing of this manuscript were performed by AI agents,
including Claude Code (Anthropic), Antigravity (Google) and ChatGPT (OpenAI), under the
audited vibe coding approach cited in the Methods~\citep{meyer2026auditedvibe}. The
author directed the process, specified the experiments and the protocol decisions,
commissioned and reviewed the independent pre-submission audit and its resolution,
reviewed the results and the text, and takes full responsibility for the finished
product.

\section*{Acknowledgements}

Computation was performed on the Cedars-Sinai high-performance computing cluster.

\bibliographystyle{unsrtnat}
\bibliography{references}

\clearpage
\appendix
\section{Supplementary tables}
\setcounter{table}{0}
\renewcommand{\thetable}{S\arabic{table}}

\begin{table}[H]
  \centering
  \caption{\textbf{Studies in the atlas and their donors.} Each accession contributes
  one tissue. Donor accessions for ENCODE studies were recovered from the experiment
  records (biosample donor); GEO studies carry no cross-series donor identifiers and are
  treated as one unit each. Counts are nuclei in the object before quality control
  (133{,}549 total).}
  \label{tab:studies}
  \setlength{\tabcolsep}{3pt}\scriptsize
  \begin{tabular}{lllrl}
    \toprule
    Accession & Tissue & Donor / sample unit & Nuclei & Notes \\
    \midrule
    ENCSR882CKY & Ventricle & ENCDO184BIH & 11{,}995 & left ventricle, male \\
    ENCSR618WVK & Liver & ENCDO757VPQ & 11{,}542 & left lobe, male \\
    ENCSR925IHI & Colon & ENCDO856ZOJ & 10{,}811 & female adult; shared donor \\
    ENCSR449VCG & Adrenal & ENCDO856ZOJ & 10{,}211 & same donor as colon, fallopian tube \\
    GSM5828478 / GSM5828491 & Bone marrow & GEO donor 15078 & 8{,}120 & healthy BMMC, age 34 \\
    GSM7661371 / GSM7661375 & Ileum & GSE238275 Line\#2 D1 & 7{,}533 & engineered inducible cell line, not primary tissue \\
    ENCSR923FKN & Atrium & ENCDO392CRK & 7{,}225 & left atrium, male \\
    ENCSR420IUS & Fallopian tube & ENCDO856ZOJ & 6{,}800 & same donor as colon, adrenal \\
    GSM5866076 / GSM5866084 & Retina & GSE196235 rep6 & 6{,}665 & adult retina \\
    GSM5884742 / GSM5884743 & Skin & GSE195452 Ctrl & 6{,}254 & forearm skin, CD90$^+$ selected \\
    GSM7096628 / GSM7096627 & Islet & GSE199636 islet 3 & 6{,}073 & primary islets \\
    ENCSR775AZP & Muscle & ENCDO575WHY & 5{,}877 & psoas, female; shared donor \\
    GSM5065527 / GSM5065528 & PBMC & GEO stimulated PBMC & 5{,}845 & 16 h IL-2 / anti-CD3/CD28 stimulation \\
    GSM7321076 / GSM7321077 & Kidney cortex & GSE232222 091422Nx & 5{,}651 & chronic kidney disease nephrectomy, male, 91 y \\
    ENCSR316WAS & Pancreas & ENCDO575EGL & 5{,}376 & female child \\
    ENCSR222ZFD & Ovary & ENCDO411EVD & 5{,}005 & female, 46 y \\
    ENCSR128ZLB & Lung & ENCDO528BHB & 4{,}947 & female, 47 y \\
    ENCSR871JTA & Bile duct & ENCDO575WHY & 4{,}269 & same donor as muscle \\
    GSM5777208 / GSM5777210 & Gray matter & GSE193240 PD003 & 3{,}350 & control brain \\
    \bottomrule
  \end{tabular}
\end{table}

\begin{table}[H]
  \centering
  \small
  \caption{\textbf{Configuration flags of the five \OT{} arms.} Everything else
  (dimensions, optimiser, schedule, batch size, peak cap, seeds, GPU) is identical.}
  \label{tab:arms}
  \footnotesize
  \begin{tabular}{llll}
    \toprule
    Arm & Encoder & Spatial prior & Decoder order \\
    \midrule
    Primary & Perceiver & chromosome mask + distance & random permutation \\
    No spatial prior & Perceiver & none & random permutation \\
    Chromosome mask only & Perceiver & chromosome mask & random permutation \\
    Genomic order & Perceiver & chromosome mask + distance & chromosome, then TSS \\
    Full attention & standard self-attention & none & random permutation \\
    \bottomrule
  \end{tabular}
\end{table}

\begin{table}[H]
  \centering
  \small
  \caption{\textbf{Hyperparameters as run.} Train time is the mean per fold as timed by
  each runner (Methods state what each timer covers).}
  \label{tab:hparams}
  \setlength{\tabcolsep}{3pt}\scriptsize
  \begin{tabular}{lccc}
    \toprule
    & \OT{} & scButterfly-B & Cisformer (support-free) \\
    \midrule
    Width & 256 & 128 (translator) & 280 \\
    Encoder & 2 cross + 2 self (256 latents) & dual VAE, chrom.-split & digit embedding, no layers \\
    Decoder layers & 2 & -- & 4 \\
    Heads & 4 & -- & 7 \\
    Peak cap & 8{,}192 & all 50{,}000 & 8{,}192 \\
    Native target & LogNormalize value & panel-normalised $\log(1+x)$ & 8 classes relative to cell max \\
    Output & gene token + Gaussian $\mu, \log\sigma^2$ & dense $3{,}185$ & 8 classes $\times$ 3{,}185 \\
    Optimiser & AdamW $3\times10^{-4}$, wd 0.01 & Adam $10^{-3}$ / SGD $5\times10^{-3}$ & Adam $5\times10^{-4}$, AMSGrad \\
    Schedule & 3\% warm-up, cosine & -- & $\times 0.6$ every 4 epochs \\
    Batch & 32 & 64 & 8 \\
    Max epochs & 100 & 30 + 30 + 60 & 30 \\
    Early stop & val loss, patience 5 & val recon.\ loss, patience 5 & val CE, patience 5 \\
    Effective seed & $7000 + 10r + d$ & 19193 (package reset) & $7000 + 10r + d$ \\
    Mixed precision & yes & package default & yes \\
    Parameters & 19.8 M (18.1 M full attn.) & package default & -- \\
    Host memory & 96 GB (64 GB ablations) & 256 GB & 96 GB \\
    Mean train time / fold & \trainHPrim{} h (\trainHFull{} h full attn.) & \trainHScb{} h & \trainHCis{} h \\
    \bottomrule
  \end{tabular}
\end{table}

\begin{table}[H]
  \centering
  \small
  \caption{\textbf{Executed checkpoint selection for the Cisformer runs, from job logs
  and checkpoint metadata.} ``Saves'' is the number of ``new best val\_loss \ldots
  saved'' lines in the Slurm log; the best epoch recorded in the manifest equals the
  epoch of minimum validation loss in the per-epoch history for every run, and each
  checkpoint file stores the validation loss as its selection score.}
  \label{tab:provenance}
  \footnotesize
  \begin{tabular}{lcccc}
    \toprule
    Run & Epochs & Best epoch (manifest) & Epoch of min.\ val.\ loss & Saves in log \\
    \midrule
    r0d0 & 24 & 19 & 19 & 14 \\
    r0d1 & 22 & 17 & 17 & 15 \\
    r1d0 & 22 & 17 & 17 & 14 \\
    r1d1 & 24 & 19 & 19 & 15 \\
    r2d0 & 24 & 19 & 19 & 17 \\
    r2d1 & 25 & 20 & 20 & 17 \\
    r3d0 & 23 & 18 & 18 & 13 \\
    r3d1 & 25 & 20 & 20 & 16 \\
    r4d0 & 24 & 19 & 19 & 16 \\
    r4d1 & 22 & 17 & 17 & 16 \\
    \bottomrule
  \end{tabular}
\end{table}

\end{document}

%% file: generated/numbers.tex
\providecommand{\nCellsTestHalf}{53{,}506}
\providecommand{\nCellsExternal}{18{,}947}
\providecommand{\nPodocyte}{299}
\providecommand{\nPaneth}{345}
\providecommand{\evaluableCV}{3{,}145}
\providecommand{\evaluableExt}{3{,}050}
\providecommand{\evaluablePod}{2{,}048}
\providecommand{\evaluablePan}{1{,}425}
\providecommand{\targetNonzero}{233}
\providecommand{\pgScb}{0.369}
\providecommand{\wcScb}{0.566}

\providecommand{\emitScb}{546}
\providecommand{\condScb}{0.424}

\providecommand{\wcPanScb}{0.657}
\providecommand{\trainHScb}{0.87}
\providecommand{\pgPrim}{0.253}

\providecommand{\emitPrim}{270}
\providecommand{\condPrim}{0.347}

\providecommand{\trainHPrim}{3.27}
\providecommand{\pgNoPrior}{0.270}
\providecommand{\wcNoPrior}{0.545}

\providecommand{\pgPanCis}{0.012}

\providecommand{\trainHCis}{7.46}

\providecommand{\wcFull}{0.369}

\providecommand{\trainHFull}{6.48}

\providecommand{\wcMean}{0.314}

\providecommand{\wcPodMean}{0.351}

\providecommand{\pgOracle}{0.996}
\providecommand{\wcOracle}{0.999}

\providecommand{\pgFullSD}{0.067}
\providecommand{\fullSDratio}{12}
\providecommand{\speedRatioPrimary}{3.8}

\providecommand{\gapWcPercMin}{0.02}
\providecommand{\gapWcPercMax}{0.04}
\providecommand{\gapPgMin}{0.10}
\providecommand{\gapPgMax}{0.24}
\providecommand{\gapWcOtherMin}{0.18}
\providecommand{\gapWcOtherMax}{0.20}
\providecommand{\pgPodTfMin}{0.02}
\providecommand{\pgPodTfMax}{0.07}
\providecommand{\pgPodBest}{0.117}
\providecommand{\extMaxDiff}{0.011}
\providecommand{\nNegPaneth}{4}
\providecommand{\nBelowMeanPodocyte}{4}

\providecommand{\repsGenomicSpatialPriorPg}{5/5}

\providecommand{\repsOutputOrderPg}{2/5}
\providecommand{\repsOutputOrderWc}{3/5}

\providecommand{\adeltaGenomicSpatialPriorPg}{0.0096}

\providecommand{\pholmGenomicSpatialPriorPg}{1.2 \times 10^{-4}}

\providecommand{\deltaWithinChromosomeDistanceGradientPg}{-0.0034}

\providecommand{\pholmWithinChromosomeDistanceGradientPg}{4.9 \times 10^{-4}}

\providecommand{\deltaPerceiverCompressionPg}{+0.0439}

\providecommand{\deltaOutputOrderPg}{-0.0007}

\providecommand{\ciLoOutputOrderPg}{-0.0024}
\providecommand{\ciHiOutputOrderPg}{+0.0010}

\providecommand{\permDescAblations}{an exact sign-flip test of the within-unit paired differences over the 16 donors (all $2^{16}$ sign patterns enumerated)}
\providecommand{\ratioPerceiverPg}{4.6}

\providecommand{\nBootAblations}{2000}

\providecommand{\adeltaGenomicSpatialPriorWc}{0.0233}

\providecommand{\deltaWithinChromosomeDistanceGradientWc}{-0.0055}

\providecommand{\pholmWithinChromosomeDistanceGradientWc}{0.014}

\providecommand{\ciLoWithinChromosomeDistanceGradientWc}{-0.0088}
\providecommand{\ciHiWithinChromosomeDistanceGradientWc}{-0.0023}

\providecommand{\deltaPerceiverCompressionWc}{+0.1838}

\providecommand{\deltaOutputOrderWc}{+0.0045}

\providecommand{\ciLoOutputOrderWc}{-0.0060}
\providecommand{\ciHiOutputOrderWc}{+0.0175}

\providecommand{\permDescAblations}{an exact sign-flip test of the within-unit paired differences over the 16 donors (all $2^{16}$ sign patterns enumerated)}
\providecommand{\ratioPerceiverWc}{7.9}

\providecommand{\nBootAblations}{2000}

\providecommand{\deltaScbVsOmicNoSpatialPriorPg}{+0.0749}

\providecommand{\deltaScbVsOmicFullAttentionPg}{+0.1188}

\providecommand{\permDescModelRanking}{an exact sign-flip test of the within-unit paired differences over the 16 donors (all $2^{16}$ sign patterns enumerated)}

\providecommand{\deltaScbVsOmicNoSpatialPriorWc}{+0.0268}

\providecommand{\pholmScbVsOmicNoSpatialPriorWc}{0.029}

\providecommand{\ciLoScbVsOmicNoSpatialPriorWc}{+0.0043}

\providecommand{\permDescModelRanking}{an exact sign-flip test of the within-unit paired differences over the 16 donors (all $2^{16}$ sign patterns enumerated)}

\providecommand{\whSeenMedianPod}{0.034}
\providecommand{\whSeenQlo}{0.003}
\providecommand{\whSeenQhi}{0.124}
\providecommand{\whSeenTypes}{26}
\providecommand{\whFracSeenBelowPod}{60\%}
\providecommand{\whFracSeenBelowPan}{2\%}
\providecommand{\whNullPgPod}{-0.000}
\providecommand{\whNullPgPan}{0.000}

\providecommand{\whPodAboveNull}{10}
\providecommand{\whPanAboveNull}{0}
\providecommand{\whTableReady}{1}
\providecommand{\capRuns}{50}
\providecommand{\capDeltaPgPod}{+0.0031}
\providecommand{\capDeltaPgMaxPod}{0.0072}

\providecommand{\capHitShortPod}{63}
\providecommand{\capHitLongPod}{0}
\providecommand{\capDeltaPgPan}{+0.0000}

\providecommand{\capDeltaPgTest}{+0.0108}

\providecommand{\nativeRunsChecked}{50}
\providecommand{\nativeCellsChanged}{0}
\providecommand{\nativeEntriesChanged}{0}
\providecommand{\nativeCapHitPct}{22.92}
\providecommand{\nativeDupsPerCell}{0.07}

%% file: generated/table_ablations.tex
\multirow{2}{*}{Genomic prior (primary vs.\ none)}
  & $\rhoG$ & $-0.0096$ & $3.1 \times 10^{-5}$ & $1.2 \times 10^{-4}$ & $[-0.0124, -0.0069]$ & 5/5 \\
  & $\rhoC$ & $-0.0233$ & $3.1 \times 10^{-5}$ & $1.2 \times 10^{-4}$ & $[-0.0356, -0.0145]$ & 5/5 \\
\addlinespace
\multirow{2}{*}{Distance gradient (primary vs.\ mask)}
  & $\rhoG$ & $-0.0034$ & $2.4 \times 10^{-4}$ & $4.9 \times 10^{-4}$ & $[-0.0048, -0.0020]$ & 5/5 \\
  & $\rhoC$ & $-0.0055$ & $0.007$ & $0.014$ & $[-0.0088, -0.0023]$ & 4/5 \\
\addlinespace
\multirow{2}{*}{Perceiver encoder (none vs.\ full attention)}
  & $\rhoG$ & $+0.0439$ & $3.1 \times 10^{-5}$ & $1.2 \times 10^{-4}$ & $[+0.0326, +0.0566]$ & 5/5 \\
  & $\rhoC$ & $+0.1838$ & $3.1 \times 10^{-5}$ & $1.2 \times 10^{-4}$ & $[+0.1478, +0.2244]$ & 5/5 \\
\addlinespace
\multirow{2}{*}{Output order (random vs.\ genomic)}
  & $\rhoG$ & $-0.0007$ & $0.487$ & $0.487$ & $[-0.0024, +0.0010]$ & 2/5 \\
  & $\rhoC$ & $+0.0045$ & $0.513$ & $0.513$ & $[-0.0060, +0.0175]$ & 3/5%

%% file: generated/table_benchmark.tex
\textit{Constant-oracle diagnostic} & \textit{null} & \textit{0.996} & \textit{0.999} & \textit{1.000} & \textit{233} & -- \\
scButterfly-B & AE & $\mathbf{0.369 \pm 0.003}$ & $\mathbf{0.566 \pm 0.005}$ & 0.871 & 546 & 0.87 \\
\OT{}, no spatial prior & TF & $0.270 \pm 0.005$ & $0.545 \pm 0.005$ & 0.735 & 273 & 2.88 \\
\OT{}, chromosome mask only & TF & $0.259 \pm 0.005$ & $0.530 \pm 0.005$ & 0.741 & 273 & 2.49 \\
\OT{}, genomic order & TF & $0.253 \pm 0.005$ & $0.527 \pm 0.006$ & 0.740 & 274 & 3.46 \\
\OT{} (primary) & TF & $0.253 \pm 0.006$ & $0.528 \pm 0.006$ & 0.731 & 270 & 3.27 \\
Cisformer, support-free & TF & $0.226 \pm 0.002$ & $0.384 \pm 0.001$ & 1.000 & 3{,}185 & 7.46 \\
\OT{}, full attention & TF & $0.134 \pm 0.067$ & $0.369 \pm 0.091$ & 0.684 & 303 & 6.48 \\
\quad Cisformer, unified sampling & TF & $0.226 \pm 0.002$ & $0.384 \pm 0.001$ & 1.000 & 3{,}185 & -- \\
\textit{Training-mean reference} & \textit{null} & \textit{0.000} & \textit{0.314} & \textit{0.000} & \textit{3{,}145} & --%

%% file: generated/table_external.tex
\textit{Constant-oracle diagnostic} & \textit{0.995} & \textit{0.999} & \textit{1.000} \\
scButterfly-B & $\mathbf{0.380 \pm 0.003}$ & $\mathbf{0.566 \pm 0.005}$ & 0.897 \\
\OT{}, no spatial prior & $0.279 \pm 0.006$ & $0.545 \pm 0.005$ & 0.750 \\
\OT{}, chromosome mask only & $0.266 \pm 0.005$ & $0.529 \pm 0.005$ & 0.754 \\
\OT{}, genomic order & $0.260 \pm 0.005$ & $0.527 \pm 0.006$ & 0.755 \\
\OT{} (primary) & $0.261 \pm 0.006$ & $0.528 \pm 0.006$ & 0.746 \\
Cisformer, support-free & $0.233 \pm 0.002$ & $0.384 \pm 0.001$ & 1.000 \\
\OT{}, full attention & $0.138 \pm 0.069$ & $0.369 \pm 0.091$ & 0.695 \\
\textit{Training-mean reference} & \textit{0.000} & \textit{0.314} & \textit{0.000}%

%% file: generated/table_ood.tex
\textit{Constant-oracle diagnostic} & \textit{0.979} & \textit{0.947} & \textit{0.999} & \textit{0.999} \\
scButterfly-B & $\mathbf{0.117}$ & $0.004$ & $\mathbf{0.437}$ & $\mathbf{0.657}$ \\
\OT{}, no spatial prior & $0.052$ & $-0.001$ & $0.347$ & $0.631$ \\
\OT{}, chromosome mask only & $0.053$ & $-0.001$ & $0.354$ & $0.571$ \\
\OT{}, genomic order & $0.043$ & $-0.000$ & $0.328$ & $0.477$ \\
\OT{} (primary) & $0.048$ & $-0.001$ & $0.347$ & $0.542$ \\
Cisformer, support-free & $0.069$ & $\mathbf{0.012}$ & $0.359$ & $0.373$ \\
\OT{}, full attention & $0.022$ & $0.000$ & $0.304$ & $0.256$ \\
\textit{Training-mean reference} & \textit{0.000} & \textit{0.000} & \textit{0.351} & \textit{0.281}%

%% file: generated/table_withheld_controls.tex
scButterfly-B & 0.117 & -0.000 & 10/10 & 0.102 & 0.004 & 0.000 & 7/10 & 0.100 \\
\OT{}, no spatial prior & 0.052 & 0.000 & 10/10 & 0.036 & -0.001 & -0.000 & 0/10 & 0.036 \\
\OT{}, chromosome mask only & 0.053 & -0.000 & 10/10 & 0.033 & -0.001 & 0.000 & 0/10 & 0.032 \\
\OT{} (primary) & 0.048 & -0.000 & 10/10 & 0.034 & -0.001 & 0.000 & 0/10 & 0.032 \\
\OT{}, genomic order & 0.043 & 0.000 & 10/10 & 0.033 & -0.000 & -0.000 & 0/10 & 0.032 \\
Cisformer, support-free & 0.069 & -0.000 & 10/10 & 0.055 & 0.012 & 0.000 & 10/10 & 0.053 \\
\OT{}, full attention & 0.022 & -0.000 & 10/10 & 0.008 & 0.000 & 0.000 & 0/10 & 0.008%